# Postural control adaptation and habituation during vibratory proprioceptive stimulation: an HD-EEG investigation of cortical recruitment and kinematics

Fabio Barollo, Rún Friðriksdóttir, Kyle J. Edmunds, Gunnar H. Karlsson, Halldór Á. Svansson, Mahmoud Hassan, Antonio Fratini, Hannes Petersen and Paolo Gargiulo*

***Abstract*— The objective of the present work is to measure postural kinematics and power spectral variation from HD-EEG to assess changes in cortical activity during adaptation and habituation to postural perturbation. To evoke proprioceptive postural perturbation, vibratory stimulation at 85 Hz was applied to the calf muscles of 33 subjects over four 75-second stimulation periods. Stimulation was performed according to a pseudorandom binary sequence. Vibratory impulses were synchronized to high-density electroencephalography (HD-EEG, 256 channels). Changes in absolute spectral power (ASP) were analyzed over four frequency bands (Δ: 0.5-3.5 Hz; θ: 3.5-7.5 Hz; α: 7.5-12.5 Hz; β: 12.5-30 Hz). A force platform recorded torque actuated by the feet, and normalized sway path length (SPL) was computed as a construct for postural performance during each period. SPL values indicated improvement in postural performance over the trial periods. Significant variation in absolute power values (ASP) was found in assessing postural adaptation: an increase in θ band ASP in the frontal-central region for closed-eyes trials, an increase in θ and β band ASP in the parietal region for open-eyes trials. In habituation, no significant variations in ASP were observed during closed-eyes trials, whereas an increase in θ, α, and β band ASP was observed with open eyes. Furthermore, open-eyed trials generally yielded a greater number of significant ASP differences across all bands during both adaptation and habituation, suggesting that following cortical activity during postural perturbation may be up-regulated with the availability of visual feedback. These results altogether provide deeper insight into pathological postural control failure by exploring the dynamic changes in both cortical activity and postural kinematics during adaptation and habituation to proprioceptive postural perturbation.**

## I. Introduction

HUMAN posture is a complex and naturally unstable physiological process that requires the continuous integration of compensatory mechanisms to maintain an equilibrium condition of upright stance [1], [2]. Collectively defined as 'postural control' [3], this process is dynamically mediated by regulatory feedback elicited from somatosensory, vestibular, and visual systems [4]. Exogenous disruption or nonspecific stimulation of these systems can induce postural sway [5], [6] – the magnitude and latency of which characterize the kinematics of postural control, which can be assessed by changes in force and torque actuated at the support surface of the feet, altogether known as posturography [7], [8], [9]. Postural control measurement is often employed under conditional balance perturbation, which is typically achieved via visual disturbance or proprioceptive stimulation [10]. Extant research in this regard cites postural control as a fundamental 'learned' motor skill, whose function and efficiency can be systematically improved with routine postural tasks [9] or directed training [11]. From these studies, two dimensions of postural learning have been posited: 'adaptation', defined as transient improvements in motor response to upright balance perturbation [12], and 'habituation', conversely defined by a gradual decrease in response to repeated perturbation [13].

While the response of sensorimotor systems during postural adaptation and habituation has been well-established in literature [14], [15] interrogating the commensurate role of the cerebral cortex or subcortical central nervous system (CNS) structures is a comparatively recent subject of research. In this regard, literature has extended previous knowledge on subcortical balance maintenance [16], [17] to consider the potential governing role of supratentorial information processing in the cerebral cortex [18]–[20]. The neuroimaging method of electroencephalography (EEG) has been cited for its high temporal resolution in measuring cortical activity [6], [21], [22]. In postural control research, balance perturbation has

Manuscript received MM DD, YY. Date of publication MM DD, YY.
This research was supported jointly by the Institute for Biomedical and Neural Engineering at Reykjavík University, the Department of Anatomy at the University of Iceland, and the Icelandic National Hospital (Landspítali Scientific Fund) with additional funding support from the Rannís Icelandic Research Fund. *Asterisk indicates corresponding author.*

R. Friðriksdóttir, K. J. Edmunds, G.H. Karlsson, H.Á. Svansson and *P. Gargiulo are with Reykjavík University, Reykjavík, Iceland (correspondence e-mail: paolo@ru.is).

Fabio Barollo is with Reykjavik University and the Department of Biomedical Engineering at Aston University, Birmingham, UK.
Antonio Fratini is with the Department of Biomedical Engineering at Aston University, Birmingham, UK.
M. Hassan is with LTSI- Université de Rennes, Rennes, France.
H. Petersen is with The Department of Anatomy, University of Iceland, Reykjavík, Iceland and with Akureyri Hospital, 600 Akureyri, Iceland
*(Fabio Barollo and Rún Friðriksdóttir are co-first authors.)*

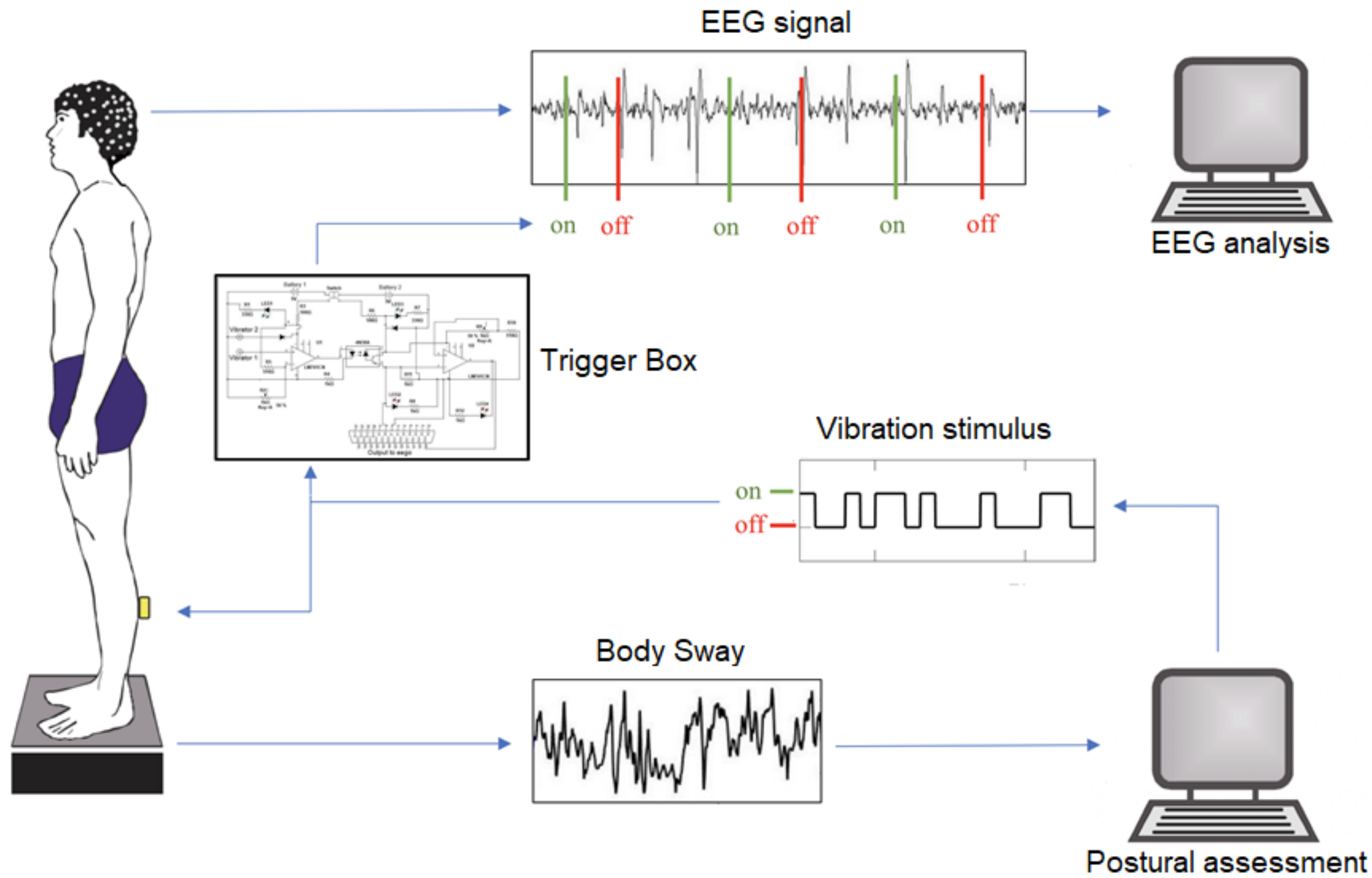

Fig. 1. Experimental set-up.

revealed scalp-level activity changes in frontal-central and frontal-parietal cortical regions, specifically within α (7.5–12.5 Hz) and θ (3.5–7.5 Hz) frequency ranges [20], [23]. Additional EEG studies report bursts of γ activity (30–80 Hz) during voluntarily anterior-posterior movements [13]. EEG activity in this regard is reported as changes in evoked time-domain event-related potentials (ERP) or as perturbation-evoked responses (PERs), e.g. N1 amplitudes and Contingent Negative Variations (CNV) [3], [6], [19], [24].

While recent evidence for the critical role of the cerebral cortex in governing postural adaptation and habituation has been reported [10], to our knowledge, cortical activity assessment during balance perturbation has never been synchronized with kinematic posturography measurements. In addition, no postural control studies report the use of 'high-density', 256-channel EEG (HD-EEG) – a methodology with superior spatial resolution to more conventional 32- or 64-channel systems. Furthermore, power spectral variation analysis from EEG data remains underreported in postural control research, despite its being a conventional method for EEG signal analysis with proven utility in cognitive and motor task studies [20], [25]. The present study aims to extend current research in this regard with the synchronized assessment of postural kinematics with power spectral variation analyses from HD-EEG to quantify changes in cortical activity during adaptation and habituation to postural perturbations using vibratory proprioceptive stimulation.

## II. Materials and Methods

### A. Experimental setup

As noted, the present experimental setup aimed at integrating posturography measurement with the assessment of cortical activity by 256-channel HD-EEG during vibratory proprioceptive stimulation. The study was reviewed and approved by the National Bioethics Committee (Vísindasiðanefnd - reference number: VSN-063) and the measurements were performed at the Icelandic Center for Neurophysiology at Reykjavik University. Thirty-three healthy volunteers (10 females and 23 males, aged 21 to 52) participated in the study. These subjects had no history of vertigo, central nervous disease, or lower extremity injury, and none of the subjects had consumed alcohol within a 24-hour period prior to their measurement. Fig. 1 illustrates the overall experimental set-up for the present work, as presented at the XV Mediterranean Conference on Medical and Biological Engineering and Computing – MEDICON 2019 [26]. Participants were instructed to maintain an upright stance during exogenous balance perturbation, evoked by the simultaneous stimulation of vibrators fastened tightly by elastic straps around the widest point of each calf.

The vibrators were designed using revolving DC-motors equipped with a 3.5 gram eccentric weight, which was contained in a cylindrical casing approximately 6 cm in length and 1 cm in diameter. Each stimulation was set to deliver vibrations of 0.1 cm in amplitude, at a frequency of 85 Hz. Stimulation was applied according to a pseudorandom binary sequence schedule (PRBS), where each shift had a random duration of 1 to 6.6 seconds.

### B. Posturography measurement

In general, maintaining a normative eased upright stance requires the symmetric distribution of body weight; when challenged, the resultant anterior-posterior or bilateral compensatory motion can be captured using a force platform to record changes in the body's center of pressure [27], [28]. For the present work, this assessment was achieved using a customized platform system developed at the Department of Solid Mechanics, Lund Institute of Technology in Sweden [3]. Anterior-posterior (ant-post) and lateral (Lat) forces actuated by the feet were recorded at six degrees of freedom with an accuracy of 0.5 N; these data were sampled at 50 Hz by a custom-made program, Postcon™, on a computer equipped with an analogue-to-digital converter. Participants were instructed to stand on a pressure plate with their arms downwardly relaxed and their feet positioned at an angle of approximately 30 degrees, open to the front, with their heels approximately three centimeters apart. Participants were asked to focus on a fixed marker point in front of them, at about 150 cm distance.

### C. HD-EEG data acquisitions

HD-EEG data were acquired using a 256-channel Ag/AgCl wet-electrode cap connected in bipolar configuration to four cascaded 64-channel amplifiers; data collection employed a standardized 10-20 system montage with EEGO software (ANT neuro, Enschede Netherlands). An additional infra-orbital electrode was used to identify any obfuscating electrooculographic (EoG) signals, and EEG data were continuously recorded at a sampling frequency of 1024 Hz. To synchronize EEG acquisition with posturography data, a custom trigger signal box was built to rectify each vibratory stimuli as a 5V TTL timing signal sent to the master amplifier. This trigger system allowed for the generation of vibratory on/off event timestamps at <1 ms latency during EEG recordings. As our previous research has identified changes in cortical activity according to the availability of visual feedback, two measurement trials were performed for each subject, beginning with open eyes (OE) and followed by closed eyes (CE). An initial quiet stance (QS) baseline phase (30 seconds) preceded each stimulation phase (300 seconds), resulting in a total duration of 330 seconds for each recording. Each stimulation phase was further subdivided into four 75-second recording periods: P1, P2, P3, and P4.

## III. Data analysis

### A. Assessing postural performance

Force platform data for both OE and CE datasets were analogously segmented into five recording periods (QS, P1, P2, P3, and P4) in order to facilitate the synchronization of postural sway data with any evoked changes in cortical activity. The normalized Sway Path Length (SPL) was computed to describe the overall postural performance of the subjects during each period.

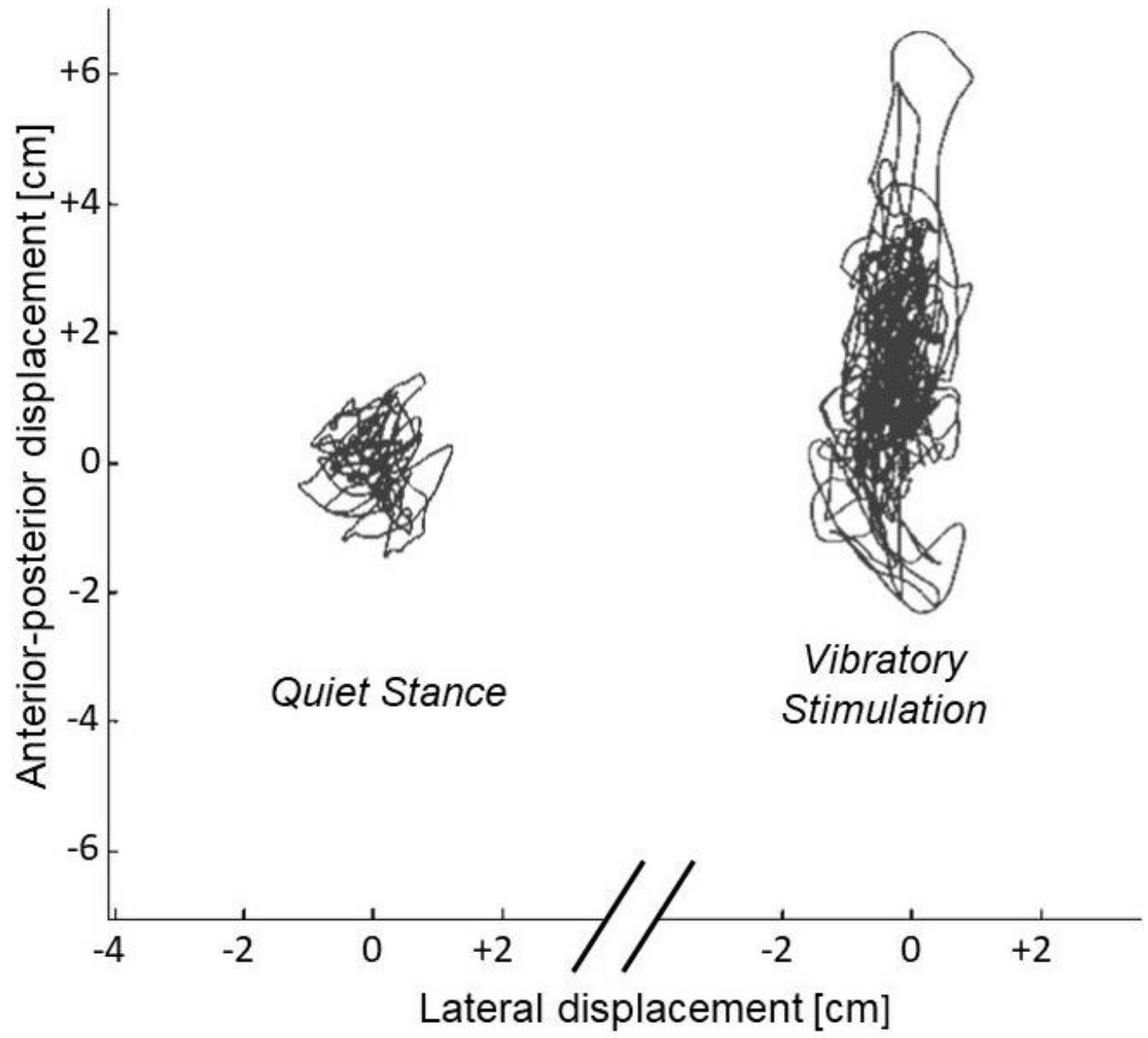

Fig. 2. Statokinesigram showing an example subject's body sway. Left: no stimulation applied. Right: randomized vibratory stimulation applied on the calves to disrupt upright stance. Anterior- posterior and lateral displacements are expressed in cm.

The center of pressure (CoP) trajectory coincides with the vertical projection on platform plane of the subject's center of mass and it is widely used in human posture studies to assess body sway [29], [30]. Torque values were extracted from the collected force platform data and used to derive ant-post and lat displacement during each trial. A graphical representation of the CoP trajectory ('statokinesigram') was then obtained by plotting the displacement along the ant-post and lat axes over time, see Fig. 2. Sway Path Length is one of several posturographic parameters that can be extracted from a postural platform and is one of the most commonly used [31]. Nevertheless, previous research has established the sensitivity of these indices to different anthropometric characteristics – particularly height and weight, where taller or heavier people appear to be more unstable [32]. As such, SPL values were normalized to each subject's height and weight to estimate the CoP trajectory on the platform according to the following formula:

$$Stabilogram_k(t) = \frac{\tau}{(0.56 \cdot h_k) \cdot (m_k \cdot 9.81)}$$

TABLE I
Normalized Sway Path Length (SPL), means, and standard deviations by experimental epoch and OE/CE condition.

| | QS | | P1 | | P2 | | P3 | | P4 | |
|---|---|---|---|---|---|---|---|---|---|---|
| Eyes | Open | Closed | Open | Closed | Open | Closed | Open | Closed | Open | Closed |
| Mean | 0.3038 | 0.4408 | 1.3624 | 1.8584 | 1.2190 | 1.7276 | 1.2312 | 1.7229 | 1.2574 | 1.6655 |
| SD | 0.0821 | 0.1392 | 0.2810 | 0.5130 | 0.2389 | 0.4447 | 0.2285 | 0.4204 | 0.2422 | 0.4285 |

Where $k$ refers to the individual participant, $0.56 \cdot h_k$ is an estimate of the center of mass's (CoM) radial distance to the platform [33] and $m_k \cdot 9.81$ is the weight applied to the participant's CoM.

### B. HD-EEG data preprocessing

Raw HD-EEG data were pre-processed using the EEGLAB Toolbox [34], first with a band-pass filter set between 0.5–80 Hz, followed by a notch filter (49.5–50.5Hz) to remove AC power line interference from each period. Three approaches to artifact rejection were performed: channel interpolation, automatic continuous artifact rejection, and principle component analysis (PCA). All channels were re-referenced to a common average.

To investigate variations in cortical activity between periods, absolute spectral power (ASP) values were obtained using fast Fourier transformation (FFT) analysis at a resolution of 0.977 Hz with a 10% Hanning window. This analysis was performed for four frequency bands: Δ (0.5-3.5 Hz), θ (3.5-7.5 Hz), α (7.5-12.5 Hz) and β (12.5-30 Hz). This analysis generated ASP values for each frequency band and period, which were extracted and exported for statistical analyses using a customized Matlab GUI (MathWorks, Inc., Natick, 158 Massachusetts, USA). From this analysis, topological maps illustrating OE and CE difference spectra were extracted for each EEG frequency band. In addition, significant changes in mean ASP at each electrode were assessed using paired heteroscedastic t-tests, which yielded topological p-value maps for each EEG waveform (with $p<0.05$ the threshold for significance). False discovery rate (FDR) and Bonferroni significance correction methods were employed and compared to address the statistical problem of multiple comparisons. Channels that resisted these multiple comparison correction methods were specifically marked in significance topologies.

## IV. Results

### A. SPL data distribution to assess adaptation and habituation periods

When describing time-varying phenomena, a distinction must be made between true adaptation resulting from control dynamic alterations and other time-dependent phenomena, such as fatigue or biomechanical alterations resultant from changes in balance conditions [3]. Because of this, we aimed to use SPL results to observe whether postural performance decreased over any of the experimental periods; with both OE and CE conditions, minimum postural sway has been reported to occur at around 150-200 seconds following incident stimuli [3], [10].

Table I shows the means and standard deviations of the normalized SPL values for the present cohort, across each of the experimental periods and conditions. Fig. 3 represents the SPL data distribution with box plots to highlight median, mean and outlier values for each period. From these results, SPL did increase in Period 4 (P4) for OE trials, but remained comparatively stable over all four periods for CE trials. To avoid any potential obfuscation from fatigue or other time-dependent variations, we therefore used the difference between Period 3 (P3) and Period 1 (P1) ASP values in our assessment of habituation.

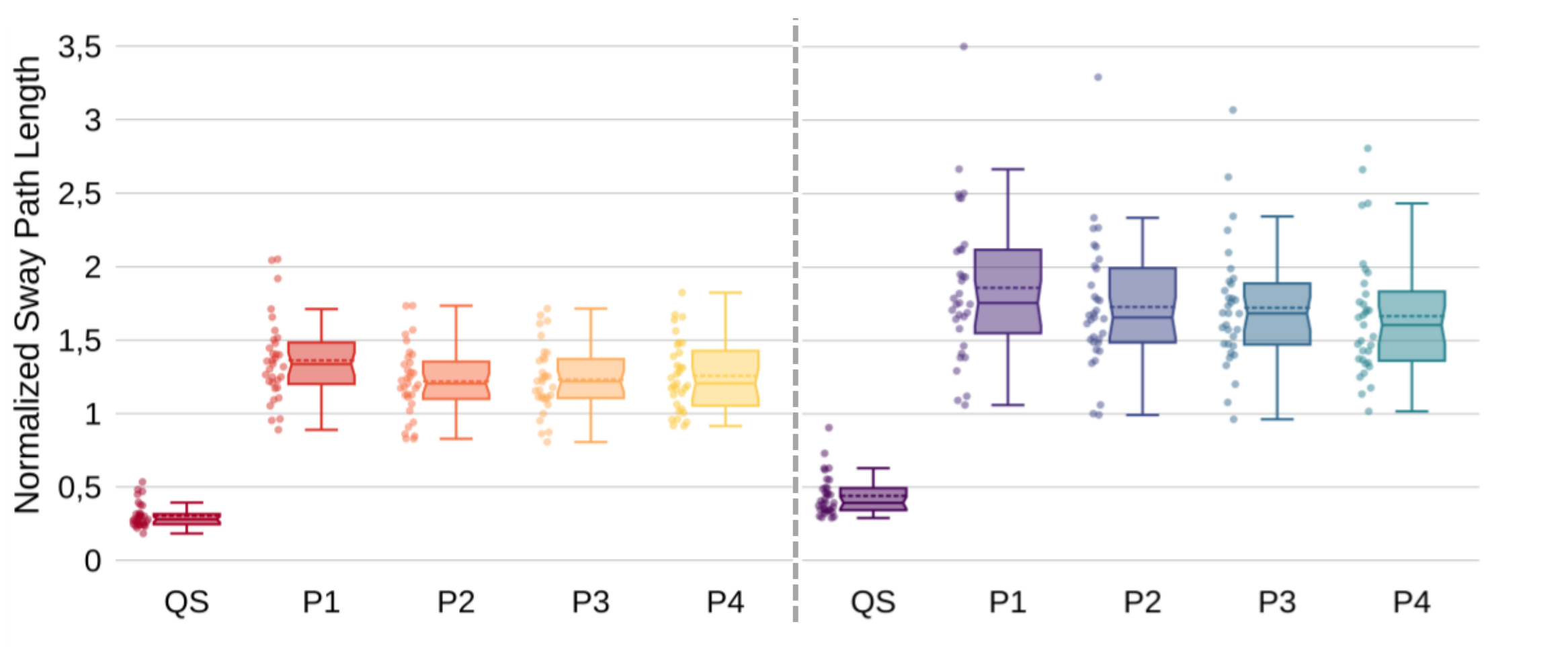

Fig. 3.Normalized Sway Path Length. Left: Open Eyes; Right: Closed Eyes.

*B. Absolute power spectra variation from HD-EEG*

Figs. 4–6 show the results from statistical analyses performed across every channel. Each of the different colored regions defines a topological map of statistical significance ($p<0.05$) corrected for multiple comparisons in accordance with the following conditions: uncorrected paired t-test (yellow), FDR correction (orange), and Bonferroni correction (red). Channels that remained significant following FDR correction are highlighted with a black 'X', while electrodes that resisted Bonferroni correction are indicated by a blue '*'. White regions are considered non-significant ($p>0.05$). In the second row, topographical maps of changes in ASP are shown solely for areas that presented a statistically significant difference between the examined periods.

***Adaptation:***

As previously mentioned, postural adaptation is shown using normalized ASP differences between P1 and QS, and Figs. 4 and 5 depict cortical maps of channels showing significant changes in CE and OE conditions, respectively. These results show an increase in ASP for both conditions, with many more significant regions present from OE trials. Generally, ASP in the θ band increased in adaptation – particularly in the frontal-central region ($p<0.05$, FDR corrected) during CE and the parietal region ($p<0.05$, FDR corrected) during OE, where ten electrodes passed the Bonferroni correction test. During OE trials, higher frequency bands (α and β) show significant activity as well. In particular, Fig. 5 shows increased activity in the parietal-occipital region in the β band.

***Habituation:***

Postural habituation is shown using normalized ASP differences between P3 and P1. Fig. 6 depicts the cortical mapping of grand mean changes in ASP solely in channels shown to have a significant difference between periods ($p<0.05$). White areas are again considered non-significant. The overall results show an increase of ASP in P3 compared to P1 for OE trials, whereas during the CE periods, no significant changes occurred. OE results specifically indicate increased activity in the θ band (temporal region), α band (parietal region), and β band (frontal region).

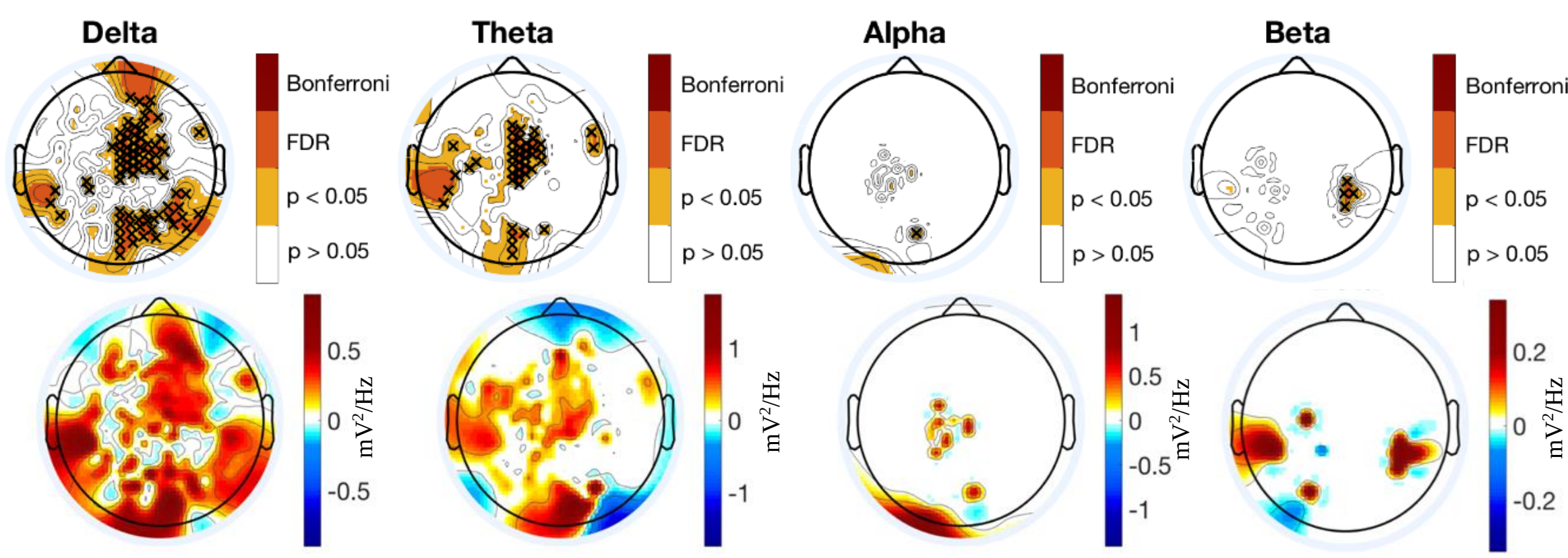

Fig. 4. Adaptation CE. Top row: statistical analysis results. Each colored region identifies topological areas that achieved statistical significance ($p<0.05$) given the following conditions: uncorrected paired t-test (yellow), FDR correction (orange), and Bonferroni correction (red). Individual electrodes that resisted FDR correction are indicated by a black 'X'. Bottom row: Topographies highlighting the average changes in ASP (between P1 and QS periods) over the whole cohort.

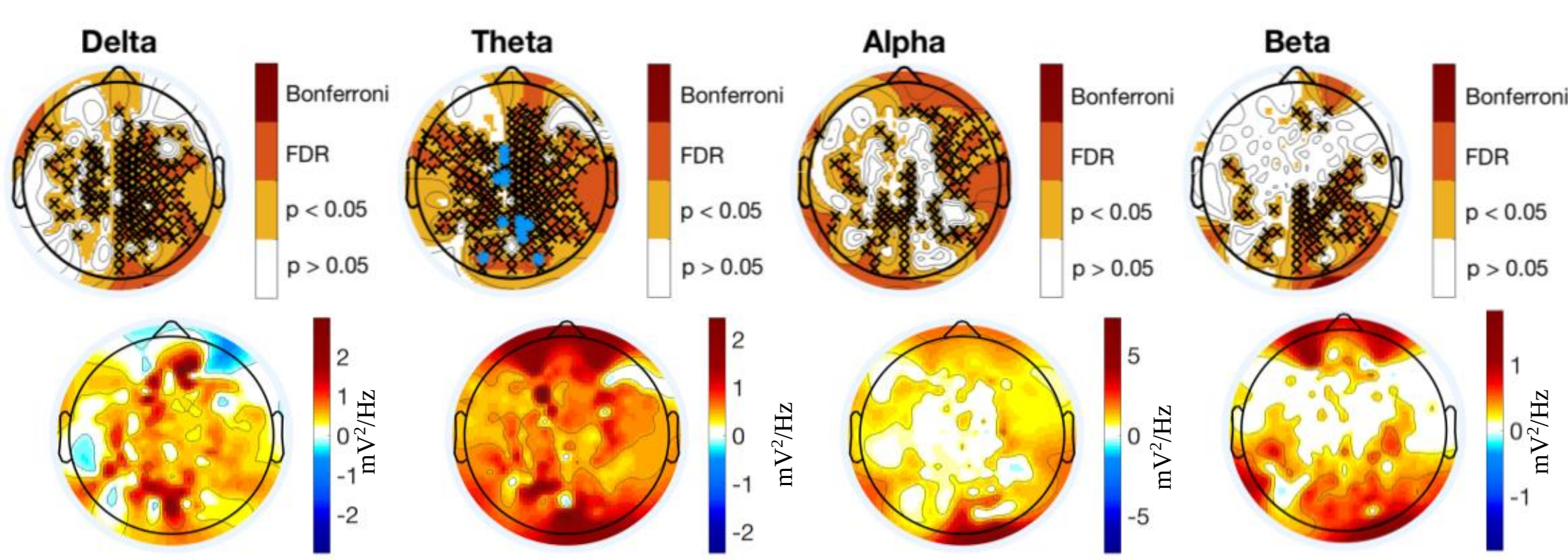

Fig. 5. Adaptation OE. Top row: statistical analysis results. Each colored region identifies topological areas that achieved statistical significance ($p<0.05$) given the following conditions: uncorrected paired t-test (yellow), FDR correction (orange), and Bonferroni correction (red). Individual electrodes that resisted FDR correction are indicated by a black 'X', while electrodes that resisted Bonferroni correction are indicated by a blue '*'.
Bottom row: Topographies highlighting the average changes in ASP (between P1 and QS periods) over the whole cohort.

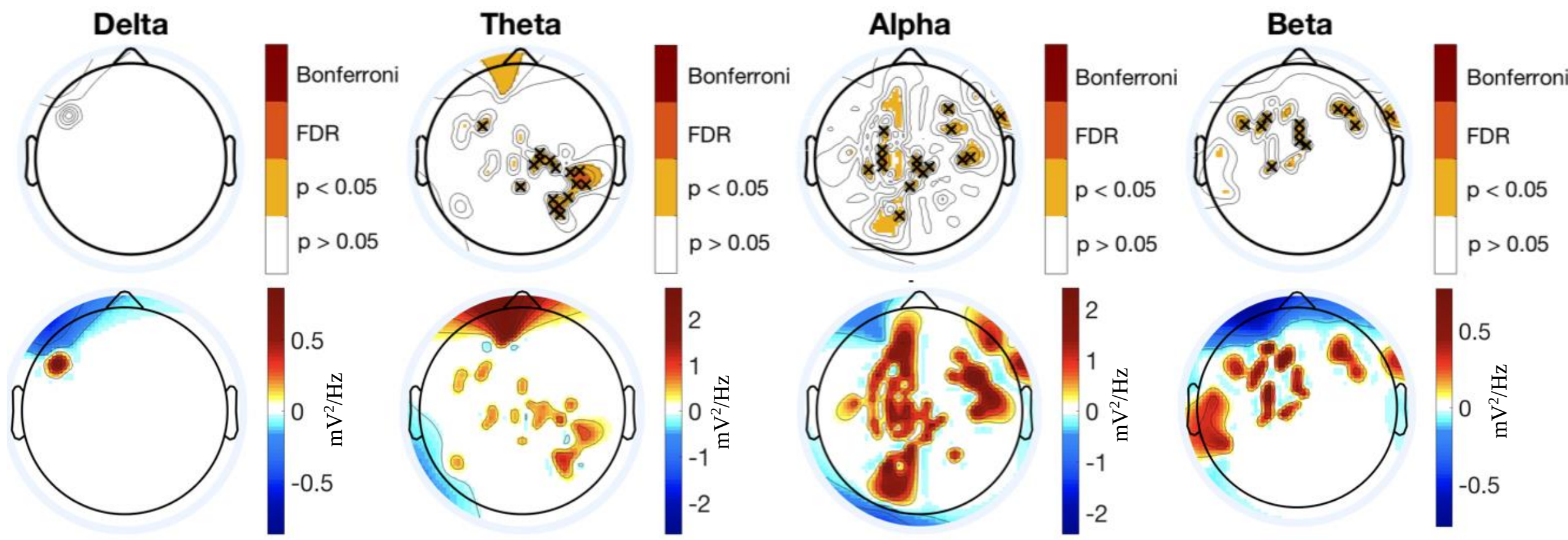

Fig. 6. Habituation OE. Top row: statistical analysis results. Each colored region identifies topological areas that achieved statistical significance ($p<0.05$) given the following conditions: uncorrected paired t-test (yellow), FDR correction (orange), and Bonferroni correction (red). Individual electrodes that resisted FDR correction are indicated by a black 'X'. Bottom row: Topographies highlighting the average changes in ASP (between P3 and P1 periods) over the whole cohort.

## V. Discussion

Power spectral variation analysis from EEG data remains underreported in postural control literature. The present study aimed to extend current research with the synchronized assessment of postural kinematics with power spectral variation analyses from HD-EEG to quantify changes in cortical activity during adaptation and habituation during a postural control task using vibratory proprioceptive stimulation.

Postural sway was recorded to obtain normalized SPL values over five experimental periods, which were then utilized to classify postural adaptation and postural habituation as differences in ASP between specific recording periods. In this regard, adaptation was defined as the difference between P1 and QS periods, in accordance with extant postural perturbation literature [3], [10], [12]. However, normalized SPL was used to define postural habituation, where P1 ASP values were subtracted from the period with the lowest mean SPL to avoid the influence of time-dependent variations such as fatigue.

There are two main limitations to note from this methodology. Firstly, while it remains possible that the onset of habituation or fatigue may differ subject-to-subject, previous studies have shown that ant-post and lat torque variations during OE and CE trials reach a consistent minimum around 150-200 seconds following incident stimuli, corresponding with P3 in the present work [3], [10]. Nevertheless, further investigation into quantitative indicators for the measurement of cortical habituation is recommended. Another methodological limitation arises from demonstrating the relationship between ASP and SPL. In this regard, we investigated whether there was any linear correlation between ASP and SPL or other metrics derived from CoP trajectory, such as the root-mean-square (RMS) of excursion in the ant-post direction. In all of these comparisons, no meaningful correlation was found, suggesting the future importance of considering alternative metrics for postural performance, such as approximate entropy or multiscale entropy, which has shown promise in linking modifications in neural involvement to responses to vibration [35].

Significant ASP differences were found across the entirety of the cortex in α and θ bands, with the exception of the prefrontal area which yielded minimal significance in the α band. Power in the θ band has been shown to increase in adaptation – particularly in the frontal-central region during CE trials and the parietal area during OE trials [20], [23], [36]. Here, our results are in accordance with the recent work by Solis-Escalante et al., who showed significant midline ASP differences in both CE and OE conditions as evidenced by the differential modulations of α and low-γ rhythms [37]. This altogether suggests that central region ASP increases during high-demand postural correction, such as balance maintenance without allowing corrective foot placement, as performed in the present study.

Furthermore, it has been suggested that increase in θ activity in the frontal-central regions is involved in error detection and processing of postural stability during balance control [22], [38]. As such, the θ band ASP differences shown here may signify the planning of corrective steps and/or the analysis of falling consequences, as indicated by our previous work on cortical functional dynamics during postural control. Relatedly, significant ASP differences in the α band may reflect an inhibition of error detection within the cingulate cortex due to habituation [10].

The present results indicate that OE trials reflect a greater number of significant differences in ASP across all bands during both adaptation and habituation. This suggests that following both acute and prolonged proprioceptive perturbation, cortical activity may be up-regulated with the availability of visual feedback. These results generally support our prior hypothesis that the visual recognition of instability may play a critical role in governing cortical processes requisite for postural control [10]. This hypothesis resulted from previous work demonstrating that all-type visual impairment is associated with an increased risk for injurious falling [39], [40]. However, in addition to the impact of vision, it is also important to note that postural control may be driven by brain network interactions rather than isolated changes in cortical activity at specific regions. Our present results support the potential importance of network dynamics, as significant differences in ASP were concurrently measured across many different cortical regions. In this context, we believe that further work investigating the reconfiguration of cortical networks during adaptation and habituation could reveal new insights about how functionally coordinated brain activity may dictate postural control. Such a model could include source-space or connectivity-based analyses, as we have previously illustrated using lower density EEG [10]. Furthermore, since the examined measures of postural performance (SPL and RMS) were not able to highlight a significant correlation between kinematics and cortical recruitment, a deeper investigation based on different postural parameters would be a fruitful area for further work.

## VI. Conclusions

Dynamic posturography is an established method for evaluating postural control. By delivering controlled disruption in the form of calf vibration, we can examine central nervous system (CNS) processing by associating body inertia and changes in upright stance [41]. This notion is true for simple upright stance but does not apply to complex postural tasks or pathological conditions, such as compensatory action during motion sickness, postural control failure from unilateral vestibular loss (UVL) [42] or cerebellar stroke [43]. In these complex postural conditions, incident adaptive and habituative processes are activated within the CNS to ensure the maintenance of upright posture and normative gait. Adding HD-EEG to dynamic posturography measurement enables the commensurate measurement of CNS activity and dynamic postural kinematics during adaptation and habituation to key postural control tasks. This invokes a deeper emphasis on the importance of further investigation into the adaptive and habitual processes implicated in CNS response to disease (i.e. UVL, cerebral and/or cerebellar diseases), which would provide key insight towards the identification of compensatory targets for clinical intervention.